\documentstyle[aps,prb,multicol]{revtex} 
\input epsf

\newcommand{\dir}{Figs}
\newcommand{\fig}[4]
{
    \noindent
    \unitlength=1mm
    \begin{picture}(#2,#3)
    \put(10,0){\leavevmode \epsfxsize=#2mm \epsffile{\dir/#1}}
    \end{picture}
    \noindent 
    #4
}
\newcommand{\rr}{ {\bf r} }
\newcommand{\ru}{ {\bf \hat{r}} }
\newcommand{\ku}{ {\bf \hat{k}} }
\newcommand{\nn}{ {\bf n} }
\newcommand{\uu}{ {\bf u} }
\newcommand{\kk}{ {\bf k} }
\newcommand{\QQ}{ {\bf Q} }
\newcommand{\UU}{ {\bf U} }
\newcommand{\II}{ {\bf I} }
\begin{document} 

\newcommand{\CCextrap}
{
\caption{
Expansion coefficient $h_{2 1 2 -1 2 0}(r)$ of the total correlation function
$h$ vs.\ $r$ in systems of size $N=8000$ (solid line), $N=4000$ (dotted line),
and $N=1000$ (dashed line). Cutoff radii were $r_{max}=11.9,9.4$, and $6.6$,
respectively. The long dashed line indicates the extrapolation towards $r \to
\infty$ (for the dataset $N=1000$). Inset shows same data vs.\ $1/r$.}
\label{fig:extrap}}

\newcommand{\CCww}
{
\caption{ 
${\cal W}_{xz}$ (a) and ${\cal W}_{yz}$ (b) surfaces for $N=4000$ (cubic box)
and $N=16 000$ (elongated box); the smaller, finer spaced grids correspond to
the larger system. The fits (dotted lines) coincide almost perfectly 
with the data (solid lines).
}
\label{fig:ww}}

\newcommand{\CCcii}
{
\caption{
Weighted sum of the DCF expansion coefficients $C_{ii}(k)$ as defined in
Eqn.~(\protect\ref{eq:cii}) vs.\ $k^2$ for different system sizes $N$
(unconstrained director, evaluated using coefficients up to $l_{max}=6$). 
The points at $k=0$ are taken from Eqn.~(\ref{eq:cond_2}). 
The initial slopes give the elastic constants $K_{ii}$. Thick solid lines 
indicate corresponding fits for the $N=4000$ system.}\label{fig:cii}}

\newcommand{\CCtaba}
{
\bigskip
\begin{tabular}{|c|ccc|@{$\:$}}
 System &  \multicolumn{3}{c|}{Order tensor fluctuations} \\
  size & 
 $\langle K_{11}\rangle$ & $\langle K_{22}\rangle$ & $\langle K_{33}\rangle$
  \\ \hline
 4000  & 0.53 $\pm$ 0.01 & 0.30 $\pm$ 0.01 & 1.60 $\pm$ 0.01  \\
 16000 & 0.53 $\pm$ 0.01 & 0.30 $\pm$ 0.01 & 1.59 $\pm$ 0.01 \\
\end{tabular}
\bigskip
\caption{\label{tab:el1}
Elastic constants from the analysis of order tensor fluctuations
for systems of different size $N$. 
}
}

\newcommand{\CCtabb}
{
\bigskip
\begin{tabular}{|c|lccc|@{$\:$}}
 System &  \multicolumn{4}{c| }{Direct correlation function } \\
  size & $l_{\mbox{\tiny max}}$ & 
  $\langle K_{11}\rangle$ & $\langle K_{22}\rangle$ & $\langle K_{33}\rangle$
    \\ \hline
1000 & 8 & 0.55 $\pm$ 0.02 & 0.35 $\pm$ 0.03 & 1.56 $\pm$ 0.04 \\
     & 6 & 0.51 $\pm$ 0.02 & 0.34 $\pm$ 0.03 & 1.52 $\pm$ 0.04 \\
     & 4 & 0.53 $\pm$ 0.03 & 0.23 $\pm$ 0.02 & 1.32 $\pm$ 0.04 \\
     & 2 & 0.51 $\pm$ 0.01 & 0.20 $\pm$ 0.01 & 1.56 $\pm$ 0.04 \\
\hline
4000 & 6 & 0.51 $\pm$ 0.02 & 0.31 $\pm$ 0.01 & 1.51 $\pm$ 0.03 \\
     & 4 & 0.65 $\pm$ 0.02 & 0.27 $\pm$ 0.02 & 1.23 $\pm$ 0.03 \\
     & 2 & 0.53 $\pm$ 0.01 & 0.22 $\pm$ 0.01 & 1.46 $\pm$ 0.03 \\
\hline
*4000& 6 & 0.52 $\pm$ 0.02 & 0.31 $\pm$ 0.01 & 1.51 $\pm$ 0.03 \\
     & 4 & 0.65 $\pm$ 0.02 & 0.27 $\pm$ 0.02 & 1.24 $\pm$ 0.04 \\
     & 2 & 0.53 $\pm$ 0.01 & 0.22 $\pm$ 0.01 & 1.48 $\pm$ 0.03 \\
\hline
8000 & 6 & 0.51 $\pm$ 0.02 & 0.33 $\pm$ 0.02 & 1.48 $\pm$ 0.03 \\
     & 4 & 0.61 $\pm$ 0.01 & 0.29 $\pm$ 0.02 & 1.25 $\pm$ 0.04 \\
     & 2 & 0.54 $\pm$ 0.01 & 0.23 $\pm$ 0.01 & 1.47 $\pm$ 0.04 \\
\end{tabular}
\bigskip
\caption{\label{tab:el2}
Elastic constants from the DCF method for systems of different 
size $N$. ($*$) marks a system that has been simulated with
a director constraint. Results are shown for different choices 
of the cutoff value $l_{max}$ in the spherical harmonics expansion 
of the pair distribution function $\rho^{(2)}$. 
See section \protect\ref{sec:results} for details.
}
}
%
%
\title{Elastic constants from direct correlation functions in 
nematic liquid crystals: a computer simulation study}

\author{Nguyen Hoang Phuong, Guido Germano, and Friederike Schmid}

\address{Fakult\"at f\"ur Physik, Universit\"at Bielefeld, 
         33615 Bielefeld, Germany}

\setcounter{page}{1}
\maketitle 

\begin{abstract}

Density functional theories such as the Poniewierski-Stecki theory relate the
elastic properties of nematic liquid crystals with their local liquid
structure, i.e., with the direct correlation function (DCF) of the particles.
We propose a way to determine the DCF in the nematic state from simulations
without any approximations, taking into account the dependence of pair
correlations on the orientation of the director explicitly. Using this scheme,
we evaluate the Frank elastic constants $K_{11}$, $K_{22}$ and $K_{33}$ in a
system of soft ellipsoids. The values are in good agreement with those
obtained directly from an analysis of order fluctuations. Our method thus
establishes a reliable way to calculate elastic constants from pair
distributions in computer simulations.

\end{abstract}
\begin{center}
PACS numbers: 61.20.Ja, 61.30.Cz, 83.10.Rs
\end{center}

%
%

\section{Introduction} 
\label{sec:intro}
\begin{multicols}{2}

Nematic liquid crystals are fluids of anisotropic particles, which are aligned
preferentially along one direction\cite{degennes,chandrasekhar}. Their
orientation is characterized by a director $\nn$ of unit length, with
physically identical states $\nn$ and $-\nn$. Since the long range 
orientational order breaks a continuous symmetry, the isotropy of space, 
there exist soft fluctuation modes --- spatial variations of the director 
$\nn(\rr)$ --- which cost no energy in the infinite wavelength limit 
(i.e., the limit where $\nn$ is rotated uniformly) and are otherwise
penalized by elastic restoring forces~\cite{chaikin,forster}. For symmetry 
reasons, the latter depend on only three material parameters at large finite 
wavelengths~\cite{degennes,chandrasekhar,chaikin,forster,oseen,frank}. 
They are described by an elastic free energy functional~\cite{frank}
\begin{eqnarray}
\label{eq:frank}
{\cal F} \{ \nn (\rr) \} &=& \frac{1}{2}\int d \rr \Big\{
K_{11} [ \nabla \cdot \nn ]^2 +
\nonumber \\ && 
K_{22} [ \nn \cdot (\nabla \times \nn) ]^2 + 
K_{33} [ \nn \times (\nabla \times \nn) ]^2 \Big\}, 
\end{eqnarray}
which has three contributions: the splay, twist and bend modes.
The parameters $K_{\alpha \alpha}$ ($\alpha = 1,2,3$), called Frank elastic
constants, control almost exclusively the structure and the properties of
nematic liquid crystals at mesoscopic length scales. Expressions that relate
them to the microscopic properties of liquid crystals are thus clearly of 
interest. 

Several microscopic approaches have been proposed and employed 
in the past~\cite{lubensky}$^-$\cite{holovko}. Poniewierski and 
Stecki~\cite{poniewierski1} have used the density
functional formalism~\cite{evans} to derive a set of 
equations which connects the elastic constants with the direct
pair correlation function (DCF), one of the central 
quantities in liquid state theories~\cite{hansen,gray}. 
In a coordinate frame where the $z$-axis points along the director $\nn$,
the equations read 
\begin{eqnarray}
\label{eq:ps1}
K_{11} &=& \frac{k_BT}{2}\int r_x^2 \: c(\rr,\uu_1,\uu_2) 
\nonumber \\ && \times
\rho^{(1)}{}'(u_{1z}) \rho^{(1)}{}'(u_{2z}) 
u_{1x}\: u_{2x}\: d\rr \: d\uu_1 \: d\uu_2, \\
\label{eq:ps2}
K_{22} &=& \frac{k_BT}{2}\int r_x^2 \: c(\rr,\uu_1,\uu_2)
\nonumber\\ && \times
\rho^{(1)}{}'(u_{1z}) \rho^{(1)}{}'(u_{2z}) 
u_{1y}\:u_{2y}\: d\rr \: d\uu_1 \:d\uu_2, \\
\label{eq:ps3}
K_{33} &=& \frac{k_BT}{2}\int r_z^2 \: c(\rr,\uu_1,\uu_2)
\nonumber\\ && \times
\rho^{(1)}{}'(u_{1z}) \rho^{(1)}{}'(u_{2z}) 
u_{1x}\: u_{2x}\: d\rr\: d\uu_1 \:d\uu_2,
\end{eqnarray}
where the vector $\rr$ connects the centers of mass of two molecules 1 and 2, 
$\uu_1$, $\uu_2$ are unit vectors along the molecule axes, 
$c(\rr,\uu_1,\uu_2)$ denotes the DCF in the nematic liquid,
and $\rho^{(1)}{}'(u_z)$ is the derivative of the one-particle
distribution function with respect to $u_z$. 
The integrals $\int \!d \rr$ run over all space, 
and $\int \! d \uu$ over the full solid angle, $T$ is the
temperature, and $k_B$ the Boltzmann constant.

Equations of the form (\ref{eq:ps1})$^-$(\ref{eq:ps3})
have later been rederived~\cite{poniewierski2}$^-$\cite{longa1}
and applied in theories~\cite{poniewierski3}$^-$\cite{zakharov4} and 
simulations~\cite{stelzer1,stelzer2,stelzer3,zakharov3}
to study elastic constants in nematic liquid crystals~\cite{footnote1}.
The main difficulty with the Poniewierski-Stecki equations is that they
depend on the DCF in the nematic phase, which is not known. 
Theories have resorted to approximations, e.g., they use a
DCF from an effectively isotropic reference 
state~\cite{singh1}$^-$\cite{paulo}, 
or from a state with perfectly aligned 
particles~\cite{osipov1,osipov2,somoza2}. 
Simulation studies~\cite{stelzer1,stelzer2,stelzer3,zakharov3}
have neglected the explicit angular dependence of the pair correlation
functions on the orientation of the director.
Longa et al.\ have recently pointed out that this approximation may
not be adequate in nematic liquid crystals\cite{longa}.

Alternatively, the elastic constants can also be determined directly 
from the long-wavelength fluctuations of the order tensor in Fourier space
\begin{equation}
\label{eq:op}
\QQ(\kk) = \frac{V}{N} \sum_{i=1}^{N}
(\frac{3}{2} \uu_i \otimes \uu_i - \frac{1}{2} {\II} ) \: 
\exp(i \kk \cdot \rr_i) 
\end{equation}
where the sum runs over all particles $i$ in the system, $\II$ denotes the
unit matrix and $\otimes$ the dyadic product of two vectors. The largest 
eigenvalue of the $3 \times 3$ matrix $\QQ$ at zero wavevector
($\QQ(\kk)|_{\kk=0}$) is the nematic order parameter $V P_2$,
and the corresponding eigenvector is the director $\nn$ of the nematic liquid. 

In a reference frame where the $z$-axis points along $\nn$ and the $y$-axis is
perpendicular to $\kk$, the order tensor fluctuations have the limiting
long-wavelength behavior\cite{forster}
\begin{eqnarray}
\label{eq:qfluc1}
\langle | Q_{xz} (\kk) |^2 \rangle 
&\stackrel{k\to0}{\sim}&
\frac{9}{4} \: 
\frac{\langle P_2 \rangle^2 V k_B T} {K_{11} k_x^2 + K_{33} k_z^3} \\
\label{eq:qfluc2}
\langle | Q_{yz} (\kk) |^2 \rangle 
&\stackrel{k\to0}{\sim}&
\frac{9}{4} \: 
\frac{\langle P_2 \rangle^2 V k_B T}{K_{22} k_x^2 + K_{33} k_z^3}.
\end{eqnarray}
Provided the simulated systems are sufficiently large, the elastic constants 
can be extracted directly from Eqns.~(\ref{eq:qfluc1}),
(\ref{eq:qfluc2})~\cite{tjipto,allen1,allen2,allen3,allen4}.

Allen et al.~\cite{allen4} have used this method to study elastic
constants in a model liquid crystal, which had already been investigated 
earlier by Stelzer et al.~\cite{stelzer1} using the
Poniewierski-Stecki equations (\ref{eq:ps1})$^-$(\ref{eq:ps3}).
The results disagreed by an order of magnitude. Since the 
determination of elastic constants via Eqns.~(\ref{eq:qfluc1}) and 
(\ref{eq:qfluc2}) is straightforward, it seems reliable and the 
values calculated by Allen et al.\ are presumably accurate. 
On the other hand, Stelzer et al.~\cite{stelzer1} use an 
``unoriented nematic approximation'', where pair correlation functions 
are replaced by their average over all orientations of the director. 
Given the importance of the Poniewierski-Stecki equations, a clearcut 
test of the applicability of Eqns.~(\ref{eq:ps1})$^-$(\ref{eq:ps3}) in a 
nematic liquid crystal is desirable. To the knowledge of the present authors, 
no one has yet employed the Poniewierski-Stecki equations with the exact DCF 
of a nematic state. This is presumably due to the fact that no method 
has been proposed so far which allows one to extract the full orientation 
dependent DCF from computer simulation data.

The present work attempts to remedy this situation. We propose a way to
calculate the DCF without any approximations from a spherical
harmonic expansion of the pair distribution function in a uniaxial nematic
liquid crystal. The expansion coefficients can be determined from computer
simulations in a straightforward manner\cite{gray}. A conveniently 
reformulated version of Eqns.~(\ref{eq:ps1})$^-$(\ref{eq:ps3}) then allows 
one to calculate the
Frank elastic constants $K_{11}$, $K_{22}$ and $K_{33}$ from a direct
inspection of expansion coefficients of the DCF in Fourier space. We apply
the method to a model system of soft ellipsoidal particles in the nematic
phase. For comparison, we also compute the Frank elastic constants from
the fluctuations of the order tensor, Eqns.~(\ref{eq:qfluc1}) and 
(\ref{eq:qfluc2}). We find that the values are in good agreement. Our 
results thus show that the Poniewierski-Stecki theory in combination with 
the correct DCF can be used to bridge between the microscopic properties 
of nematic liquid crystals and their mesoscopic, i.e., elastic properties.

Our paper is organized as follows. We develop the theoretical tools needed 
for our procedure in section II. Section III gives details of the 
simulation model and the simulation techniques. The results are 
presented in section IV and discussed in section V.

\section{Theoretical background} 
\label{sec:theory}

We begin by recalling some common definitions\cite{zannoni}.
Let us denote by $\rho(\uu,\rr)$ the local number density of particles 
with orientation $\uu$ at position $\rr$. In a uniaxial nematic liquid 
at equilibrium with director $\nn_0$, it is distributed according to a 
one-particle distribution function 
$\langle \rho(\uu,\rr)\rangle = \rho^{(1)}(\uu)$, 
that actually depends on $|\uu \cdot \nn_0|$ only.
The pair distribution function $\rho^{(2)}(\uu_1,\uu_2,\rr_1-\rr_2)$
gives the probability of finding a particle with the orientation $\uu_1$
at the position $\rr_1$, and another particle with orientation $\uu_2$
at $\rr_2$. Particles at infinite distance become uncorrelated, hence
$\rho^{(2)}(\uu_1,\uu_2,\rr) \stackrel{ r \to \infty}{\longrightarrow}
\rho^{(1)}(\uu_1) \rho^{(1)}(\uu_2)$. 
This motivates the definition of the so-called total correlation function 
\begin{equation}
\label{eq:TCF}
h(\uu_1,\uu_2,\rr) = \frac{\rho^{(2)}(\uu_1,\uu_2,\rr)}
{\rho^{(1)}(\uu_{1}) \rho^{(1)}(\uu_{2})} -1,
\end{equation}
which measures the total effect of a particle 1 on a particle 2.
This effect is often separated into two parts: a hypothetical ``direct'' 
effect of 1 on 2, characterized by the direct correlation function 
$c(\uu_1,\uu_2,\rr)$ and an ``indirect'' effect, where 1 is assumed to
influence other particles 3, 4, etc., which in turn affect 2.
The total correlation function is related to the DCF 
via the Ornstein-Zernike equation\cite{hansen}
\begin{eqnarray}
\label{eq:OZ}
\lefteqn{h({\uu_1,\uu_2,\rr_{12}}) = c(\uu_1,\uu_2,\rr_{12}) +} 
& \qquad &
\nonumber\\ &&
\int c(\uu_1,\uu_3,\rr_{13}) \: \rho^{(1)}(\uu_3) \:
 h(\uu_3,\uu_2,\rr_{32}) d\uu_{3} d\rr_{3}, 
\end{eqnarray}
where $\rr_{ij}$ abbreviates $\rr_i - \rr_j$.

In the framework of density functional theories, the direct correlation 
function has another interpretation as the second functional derivative 
of the excess free energy with respect to local density distortions 
$\delta \rho(\uu,\rr) = \rho(\uu,\rr)-\rho^{(1)}(\uu\cdot\nn_0)$\cite{hansen}.
To lowest order in $\delta \rho$, the expansion of the free energy 
functional about an undistorted equilibrium reference state is given by
\begin{eqnarray}
\delta^{2} {\cal F} & = &
\frac{k_BT}{2} \int \left[
\frac{\delta(\uu_1 -\uu_2)\delta({\rr_{12}})}
{\rho^{(1)}(\uu_1 \cdot \nn_0)}
- c(\uu_1,\uu_2,\rr_{12}) \right] \: 
\nonumber \\ &&
\label{eq:df2_2} 
\times \delta\rho(\uu_1,\rr_1) \: \delta\rho(\uu_2,\rr_2) \: 
d\rr_1 \: d\rr_2 \: d\uu_1 \: d\uu_2. 
\end{eqnarray} 
In systems of particles with uniaxial symmetry, further approximations 
are not needed~\cite{somoza}. However, the derivation is greatly 
simplified by the additional assumption that the relevant long-wavelength 
distortions can be expressed as local distortions of the director 
$\nn(\rr)$, and that the density distribution is otherwise at local 
equilibrium~\cite{poniewierski1} 
\begin{equation}
\label{eq:drho}
\rho(\uu,\rr) \approx \rho^{(1)}(\uu \cdot \nn(\rr)).
\end{equation}
Expanding the free energy in terms of $\delta \nn(\rr) = \nn(\rr)-\nn_0$ 
rather than $\delta \rho(\uu,\rr)$ and switching to a representation in
Fourier space, Eqn.~(\ref{eq:df2_2}) then reads
\begin{eqnarray}
\label{eq:df2_3}
\delta^{2} {\cal F}  &=& 
\frac{V k_BT}{2} \!\! \int \left[
\frac{\delta(\uu_1-\uu_2)}{\rho^{(1)}(\uu_1\cdot \nn_0)}-c(\uu_1,\uu_2,\kk)
\right] \: \\ &&\times
\rho^{(1)}{}'(\uu_1\cdot \nn_0) \: 
\rho^{(1)}{}'(\uu_2 \cdot \nn_0) \:
\nonumber\\ &&\times
[\uu_1 \cdot \delta \nn(\kk)] \:
[\uu_2 \cdot \delta \nn(-\kk)] \:
d\kk \: d\uu_1 \: d\uu_2 . \nonumber
\end{eqnarray}
This expression has to be related to Eqn.~(\ref{eq:frank}),
which has the Fourier representation
\begin{eqnarray}
\label{eq:frank_fourier}
{\cal F} \{ \nn (\kk) \} &=& \frac{1}{2}\int d \kk \Big\{
K_{11} [ \kk \cdot \nn ]^2 +
\nonumber \\ && 
K_{22} [ \nn \cdot (\kk \times \nn) ]^2 + 
K_{33} [ \nn \times (\kk \times \nn) ]^2 \Big\}. 
\end{eqnarray}
To this end, we expand the DCF $c (\uu_1,\uu_2,\kk)$ in 
Eqn.~(\ref{eq:df2_3}) in powers of $\kk$ up to second order. 
For convenience, we choose a coordinate frame such that
the $z$-axis points in the direction of $\nn_0$ (director frame).

Since a global rotation of the director $\nn$ does not change the 
free energy, the leading term $\kk = 0$ must vanish, i.e., one has
\begin{eqnarray}
\lefteqn{\int \frac{\rho^{(1)}{}'(u_z)^2}
{\rho^{(1)}(u_z)} \: u_{\alpha}^2 d \uu
= 
\int \!\! c(\uu_1,\uu_2,\kk=0) \:
\rho^{(1)}{}'(u_{1,z}) \: 
} \qquad \qquad \qquad&&
\nonumber \\ &&
\label{eq:cond}
\times 
\rho^{(1)}{}'(u_{2,z}) \: 
u_{1,\alpha} \: u_{2,\alpha} \:
d\uu_1 \: d\uu_2 
\end{eqnarray}
for $\alpha = x,y$. Eqn.~(\ref{eq:cond}) has been derived in a different 
context by Gubbins\cite{gubbins} and is quite generally valid.
For symmetry reasons, the terms linear in $\kk$ in the expansion
of (\ref{eq:df2_3}) vanish too. The quadratic terms lead to an 
expression of the form (\ref{eq:frank_fourier}), with $K_{ii}$ given by
\begin{eqnarray}
\label{eq:ps1_fourier}
K_{11} &=& -\frac{k_BT}{2}\int  
\frac{\partial^2 c(\kk,\uu_1,\uu_2) }{\partial k_x^2} \Big|_{\kk = 0}
\nonumber \\ && \times
\rho^{(1)}{}'(u_{1z}) \rho^{(1)}{}'(u_{2z}) 
u_{1x}\: u_{2x}\:  d\uu_1 \: d\uu_2, \\
\label{eq:ps2_fourier} 
K_{22} &=& -\frac{k_BT}{2}\int  
\frac{\partial^2 c(\kk,\uu_1,\uu_2) }{\partial k_x^2} \Big|_{\kk = 0}
\nonumber\\ && \times
\rho^{(1)}{}'(u_{1z}) \rho^{(1)}{}'(u_{2z}) 
u_{1y}\:u_{2y}\: d\uu_1 \:d\uu_2, \\
\label{eq:ps3_fourier}
K_{33} &=& -\frac{k_BT}{2}\int 
\frac{\partial^2 c(\kk,\uu_1,\uu_2) }{\partial k_z^2} \Big|_{\kk = 0}
\nonumber\\ && \times
\rho^{(1)}{}'(u_{1z}) \rho^{(1)}{}'(u_{2z}) 
u_{1x}\: u_{2x}\:  d\uu_1 \:d\uu_2,
\end{eqnarray}
which is the Fourier space version of the Poniewierski-Stecki 
equations (\ref{eq:ps1})--(\ref{eq:ps3}). As mentioned
above, the same result can be derived without the approximation 
(\ref{eq:drho}) for systems of particles with uniaxial 
symmetry~\cite{somoza}. Compact expressions for the correction 
terms in systems of asymmetric molecules have been given by 
Yokoyama~\cite{yokoyama}. In this paper, we shall be
concerned with uniaxially symmetric molecules only.

For practical applications, it is convenient to expand all 
orientation dependent functions in spherical harmonics $Y_{lm}(\uu)$. 
In the director frame, we obtain
\begin{equation}
\label{eq:fl}
\rho^{(1)}(\uu) = \varrho \sum_{l \: \mbox{\tiny even} } f_{l} \:Y_{l 0}(\uu),
\end{equation} 
where $\varrho$ is the total bulk number density, and
\begin{eqnarray}
F(\uu_1,\uu_2,\rr)& =&
\sum_{{{l_1,l_2,l}\atop {m_1,m_2,m}}} \!\!
F_{l_1 m_1 l_2 m_2 l m}(r) \:
\nonumber\\
\label{eq:Fl}
&& \qquad \qquad
Y_{l_{1}m_{1}}(\uu_1) \: Y_{l_{2}m_{2}} (\uu_2) \: Y_{lm}(\ru).
\\
F(\uu_1,\uu_2,\kk)& =&
\sum_{{{l_1,l_2,l}\atop {m_1,m_2,m}}} \!\!
F_{l_1 m_1 l_2 m_2 l m}(k) \:
\nonumber\\
\label{eq:Fl_fourier}
&& \qquad \qquad
Y_{l_{1}m_{1}}(\uu_1) \: Y_{l_{2}m_{2}} (\uu_2) \: Y_{lm}(\ku).
\end{eqnarray}
Here $F$ stands for any of $\rho^{(2)}$, $h$, or $c$, $\ru$ denotes the
unit vector $\rr/r$, and $\ku$ the unit vector $\kk/k$.
The symmetry of the nematic phase ensures that all
coefficients are real and only coefficients with $m+m_1+m_2=0$, and 
$l+l_1+l_2$ even, enter the expansions (\ref{eq:Fl}) and (\ref{eq:Fl_fourier}). 
If the molecules have uniaxial symmetry, every single $l_i$ has to 
be even in addition.

Next we derive matrix versions of Eqns.~(\ref{eq:TCF}) 
and (\ref{eq:OZ}). To simplify the expressions, we introduce the notation
\begin{eqnarray}
&&
\Gamma^{l \: l' l''}_{m m' m''} = 
\int d\uu \: Y_{lm}^*(\uu) Y_{l',m'}(\uu), Y_{l'',m''}(\uu) \\
\lefteqn{ = \sqrt{\frac{(2l''+1)(2l' +1)}{4\pi (2l+1)}}
C(l'' l' l;0 0 0) C(l''l'l;m''m'm), } \nonumber
\end{eqnarray}
where $C$ are the Clebsch-Gordan coefficients. 
The total correlation function $h$ can then be calculated from 
$\rho^{(2)}$ by inversion of the matrix version of Eqn.~(\ref{eq:TCF})
\begin{eqnarray}
\lefteqn{
\rho^{(2)}_{l_1 m_1 l_2 m_2 l m}(r) = 
\varrho^{2} \Big(\:
\sqrt{4 \pi} f_{l_1} f_{l_2} 
\delta_{m_1 0}\delta_{m_2 0}\delta_{l 0} \delta_{m 0} 
} \qquad
\nonumber \\&&
\label{eq:TCF_2}
+ \: 
\sum_{{l_1' l_1''}\atop{l_2',l_2''}} 
h_{l_1' m_1 l_2' m_2 l m}(r) 
f_{l_1''} f_{l_2''} 
\Gamma^{l_1 l_1' l_1''}_{m_1 m_1 0} 
\Gamma^{l_2 l_2' l_2''}_{m_2 m_2 0} \:
\Big).
\end{eqnarray}
Eqn.~(\ref{eq:TCF_2}) is a linear system of equations and can be 
solved for the coefficients of $h$ by standard numerical methods.

The Ornstein-Zernike equation (\ref{eq:OZ}) is most conveniently solved
in Fourier space $\kk$. We calculate the coefficients 
$h_{l_1 m_1 l_2 m_2 l m}(k)$ of the total correlation function
in Fourier space by using the Hankel transformation\cite{gray}
\begin{equation}
\label{eq:hankel}
h_{l_1 m_1 l_2 m_2 l m}(k) = 4\pi i^l \!\!
\int_{0}^{\infty}\!\!\!\!\!\!\!
 r^2\, j_l(kr) \, h_{l_1 m_1 l_2 m_2 l m}(r)\: dr,
\end{equation}
with the spherical Bessel functions $j_l$. 
The matrix version of the 
Ornstein-Zernike equation (\ref{eq:OZ}) in Fourier space reads 
\begin{eqnarray}
h_{l_1 m_1 l_2 m_2 l m }(k) & = &
c_{l_1 m_1 l_2 m_2 l m}(k) \nonumber \\ 
\lefteqn{
+ \varrho \sum_{{l_3 l_3' l_3'' m_3}\atop {l' m' l'' m'' l_3}}\!\!
c_{l_1 m_1 l_3 m_3 l' m'}(k) \:
h_{l_3' m_3 l_2 m_2 l'' m''}(k) \: 
} \qquad \qquad 
\nonumber\\ &&
\label{eq:OZ_2}
\times f_{l_3''} \:
\Gamma^{l l' l''}_{m m' m''}
\Gamma^{l_3 l_3' l_3''}_{m_3 m_3 0} (-1)^{m_3}. 
\end{eqnarray}
The result for the direct correlation function $c(k)$ is readily 
transformed back into real space by another Hankel transformation.
However, this is not necessary for our purpose, because the 
Poniewierski-Stecki equations assume a very handy form in Fourier space: 
the spherical harmonic representation of 
Eqns.~(\ref{eq:ps1_fourier})$^-$(\ref{eq:ps3_fourier}) reads
\begin{equation}
\label{eq:ps_2}
K_{ii} = \frac{1}{2} \frac{d^2}{dk^2} \: C_{ii}(k) \Big|_{k=0}
\qquad \mbox{for $i=1,2,3$ }
\end{equation}
with
\begin{eqnarray}
\label{eq:cii}
C_{ii}(k) &=& \frac{k_B T \varrho^2}{8\sqrt{\pi}}
\sum_{l_1 l_2}\sqrt{l_1(l_1+1)}\sqrt{l_2(l_2+1)} \: f_{l_1}\: f_{l_2} 
\nonumber \\
\times &\Big\{& \: [c_{l_1 1 l_2 -1 0 0}(k) + c_{l_1 -1 l_2 1 0 0}(k)]
\nonumber \\ && \:
 + \: v_i \frac{\sqrt{5}}{2} 
   [c_{l_1 1 l_2 -1 2 0}(k) + c_{l_1 -1 l_2 1 2 0}(k)] 
\nonumber \\ && \:
 + \: w_i \frac{\sqrt{15}}{\sqrt{8}}
   [c_{l_1 1 l_2 1 2 -2}(k) + c_{l_1 -1 l_2 -1 2 2}(k)] \: \Big\}
\end{eqnarray}
and $ (v_1,v_2,v_3) = (-1,-1,2)$, $(w_1,w_2,w_3) = (-1,1,0)$.
Deriving these equations, we have exploited the relation
$\int \!\! d\rr \: F(\rr) \: r_{\alpha}^2 = 
 - {\partial^2} F(\kk)/{\partial k_{\alpha}^2} |_{\kk = 0}$
and properties of spherical harmonics. 
Finally, Eqn.~(\ref{eq:cond}) can be rewritten as
\begin{equation}
\label{eq:cond_2}
C_{ii}(k=0) = - k_B T \: \pi
\int_{-1}^{1} du_z \: (1-u_z^2) \frac{\rho^{(1)}{}'(u_z)^2}{\rho^{(1)}(u_z)},
\end{equation}
where $C_{ii}(k)$ is defined as in Eqn.~(\ref{eq:cii}). 

\section{Model and Simulation Details} 
\label{sec:simulation}

We performed computer simulations of a system of axially symmetric
rigid particles, which interact via a simple repulsive pair potential
\begin{equation}
V_{ij}
= \left\{ \begin{array}{lcr}
4 \epsilon_0 \: (X_{ij}^{12} - X_{ij}^{6}) + \epsilon_0 & : & X_{ij}^6 > 
1/2 \\
0 & : & \mbox{otherwise}
\end{array} \right. .
\end{equation}
Here $X_{ij} = \sigma_0/(r_{ij}-\sigma_{ij}+\sigma_0)$, 
$r_{ij}$ denotes the distance between particles $i$ and $j$, 
and the shape function
\begin{eqnarray}
\lefteqn{\sigma_{ij}(\uu_i,\uu_j,\ru_{ij}) =
\sigma_0 \: \left\{ 1 - \frac{\chi}{2} \left[
\frac{(\uu_i\cdot\ru_{ij}+\uu_j\cdot\ru_{ij})^2}
     {1+\chi \uu_i\cdot\uu_j} \right. \right.
}\qquad \qquad \nonumber \\&& \left. \left. + 
\frac{(\uu_i\cdot\ru_{ij}-\uu_j\cdot\ru_{ij})^2}
     {1-\chi \uu_i\cdot \uu_j} \right] \right\}^{-1/2},
\end{eqnarray}
approximates the contact distance between two ellipsoids of
elongation 
$\kappa = \sigma_{\mbox{\tiny end-end}}/\sigma_{\mbox{\tiny side-side}}
= \sqrt{(1+\chi)/(1-\chi)}$ with orientations $\uu_i$ and $\uu_j$,
which are separated by a center-center vector
in the direction of $\ru_{ij}=\rr_{ij}/r_{ij}$~\cite{berne}. 
We use throughout scaled units defined in terms of $\epsilon_0$, 
$\sigma_0$, the particle mass $m_0$ and the Boltzmann constant $k_B$. 
We studied systems of particles with elongation $\kappa = 3$
at temperature $T=0.5$ and number density $\varrho=0.3$. 
The pressure was $P = 2.60$~\cite{footnote2}. This corresponds to 
a state well in the nematic phase: at fixed temperature $T=0.5$, 
the fluid remains nematic down to the density $\varrho=0.29$ or, 
equivalently, the pressure $P=2.35$~\cite{harald}. 
The average order parameter density in our system was 
$\langle P_2 \rangle = 0.69$ and the fourth rank parameter was 
$\langle P_4(\uu \cdot \nn_0) \rangle = 0.31$,
$P_4(x)=(35 x^4-30 x^2+3)/8 $ being the fourth Legendre polynomial. 

The pair distribution function was determined in systems of $N$ = 1000, 
4000 and 8000 particles in cubic boxes with periodic boundary conditions. 
For the $N=1000$ system we used a Monte Carlo (MC) program by
H.~Lange\cite{harald}. Trial moves picked a particle at random 
and attempted in random order either a rotation or a translation, 
with maximum step sizes chosen such that the Metropolis acceptance 
rate was roughly 30\%. The larger systems were studied with a 
massively parallel computer, using a domain decomposition molecular
dynamics (MD) program, that has been codeveloped by one of us (GG). 
These simulations were performed in the microcanonical ensemble 
using the \textsc{rattle} integrator~\cite{andersen,allentildesley} 
with time step $\Delta t = 0.003$~\cite{footnote3} and molecular moment of
inertia $I$ = 2.5. Run lengths were 8 million MC steps, one MC step 
consisting of 2$N$ trial moves, or 10 million MD steps, respectively; 
data for the pair distribution function were collected every 
1000 or 10000 steps.

The order tensor fluctuations are sampled most efficiently if the 
$\kk$-vectors in Eqns.~(\ref{eq:qfluc1}) and (\ref{eq:qfluc2}) are always on 
the same grid. They were therefore determined from independent simulations 
in an ensemble where the director $\nn_0$ was constrained to the $Z$-axis 
of the simulation box~\cite{allen4}. Thus the $xyz$ frame of 
Eqns.~(\ref{eq:qfluc1}) and (\ref{eq:qfluc2}) becomes coincident 
with the $XYZ$ frame of the simulation box. The constraint was implemented 
in the MD simulations by adding two global Lagrange multipliers to the 
integrator, so that $Q_{XZ}(0) = Q_{YZ}(0) =0$ at every time step.
Our procedure was similar to that introduced by Allen et al.~\cite{allen4},
except that we used an improved integrator~\cite{germano1}
designed in the spirit of \textsc{rattle}~\cite{andersen,allentildesley},
so that it is symplectic and fulfills the constraints exactly. 
The same integrator has already been used~\cite{germano2} to calculate 
$K_{22}$ in a Gay-Berne fluid~\cite{gay}; the value compared well with an 
estimate from a thermodynamic perturbation approach.
Here, we simulated a system of $N=4000$ particles in a cubic box
over 10 million MD steps, and a system of $N=16000$ particles in an elongated 
box with side ratios $L_X:L_Y:L_Z = 1:1:2$~\cite{footnote4} over 5
million MD steps. Data for the order tensor were collected every 200 steps. 
The largest autocorrelation times were of the order of $10^5$ MD 
steps at the lowest $k$ values and dropped rapidly below 1000 
MD steps for higher $k$. 

\end{multicols}\twocolumn
\section{Data Analysis and Results}
\label{sec:results}

We begin by presenting the results for the order tensor fluctuations. 
Following Ref.~\onlinecite{allen4}, we calculated the quantities
\begin{eqnarray}
\label{eq:ww1}
{\cal W}_{x z}(\kk) &= &
\frac{9 \langle P_2 \rangle^2 V k_B T}
{4 \langle | Q_{ x z} (\kk) |^2 \rangle }
\stackrel{k \to 0}{\sim}
K_{11} k_x^2 + K_{33} k_z^2 \\
\label{eq:ww2}
{\cal W}_{y z}(\kk) &= &
\frac{9 \langle P_2 \rangle^2 V k_B T}
{4 \langle | Q_{ y z} (\kk) |^2 \rangle }
\stackrel{k \to 0}{\sim}
K_{22} k_x^2 + K_{33} k_z^2,
\end{eqnarray}
where the frame is chosen such that $\kk$ lies in the $xz$-plane 
(cf.~Eqns.~(\ref{eq:qfluc1}) and (\ref{eq:qfluc2})). More specifically,
we evaluated the order tensor in Fourier space $\QQ(\kk)$ on a $\kk$-grid
with $6 \times 6 \times 6$ grid points in the small system ($N=4000$), and 
$6 \times 6 \times 12$ grid points in the large system ($N=16000$). Then we 
applied a rotation 
$\QQ^{(xyz)}(\kk) = \UU(\kk) \QQ^{(XYZ)}(\kk) \UU^{\sf T}(\kk)$
into the desired coordinate frame such that $k_y$ = 0, 
and calculated the
averages $\langle |\QQ_{ \alpha z} (\kk) |^2 \rangle$ and the
${\cal W}_{\alpha z}(\kk)$-surface in that frame.
Because of the constraint on $\nn_0$, $\UU(\kk)$ is a constant
throughout the run. 

In the high wavelength-limit $k\to \infty$,  
${\cal W}_{\alpha z}(\kk)$ ($\alpha$=1,2) takes the value~\cite{allen4} 
\begin{equation}
{\cal W}_{\alpha z}(\kk) 
\stackrel{k \to \infty}{\longrightarrow}
\frac{\langle P_2 \rangle^2 \rho k_B T}
{ (\langle P_2 \rangle /21 - 4 \langle P_4 \rangle/35 + 1/15}.
\end{equation}

\begin{figure}[t]

\vspace*{-0.5cm}
\noindent
\hspace*{-1cm}
\fig{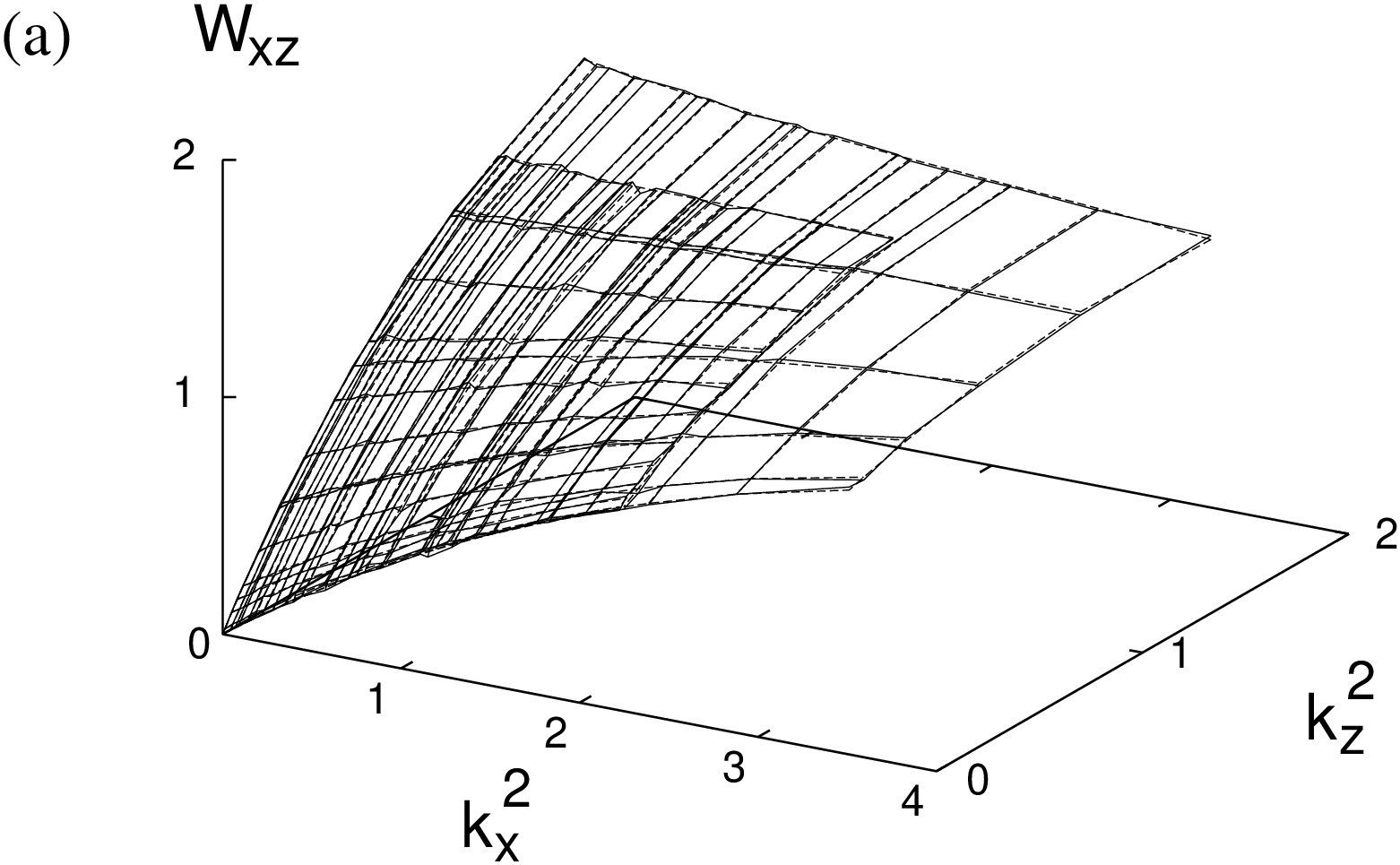}{63}{60}{
\vspace*{-1cm}
}

\noindent
\hspace*{-1cm}
\fig{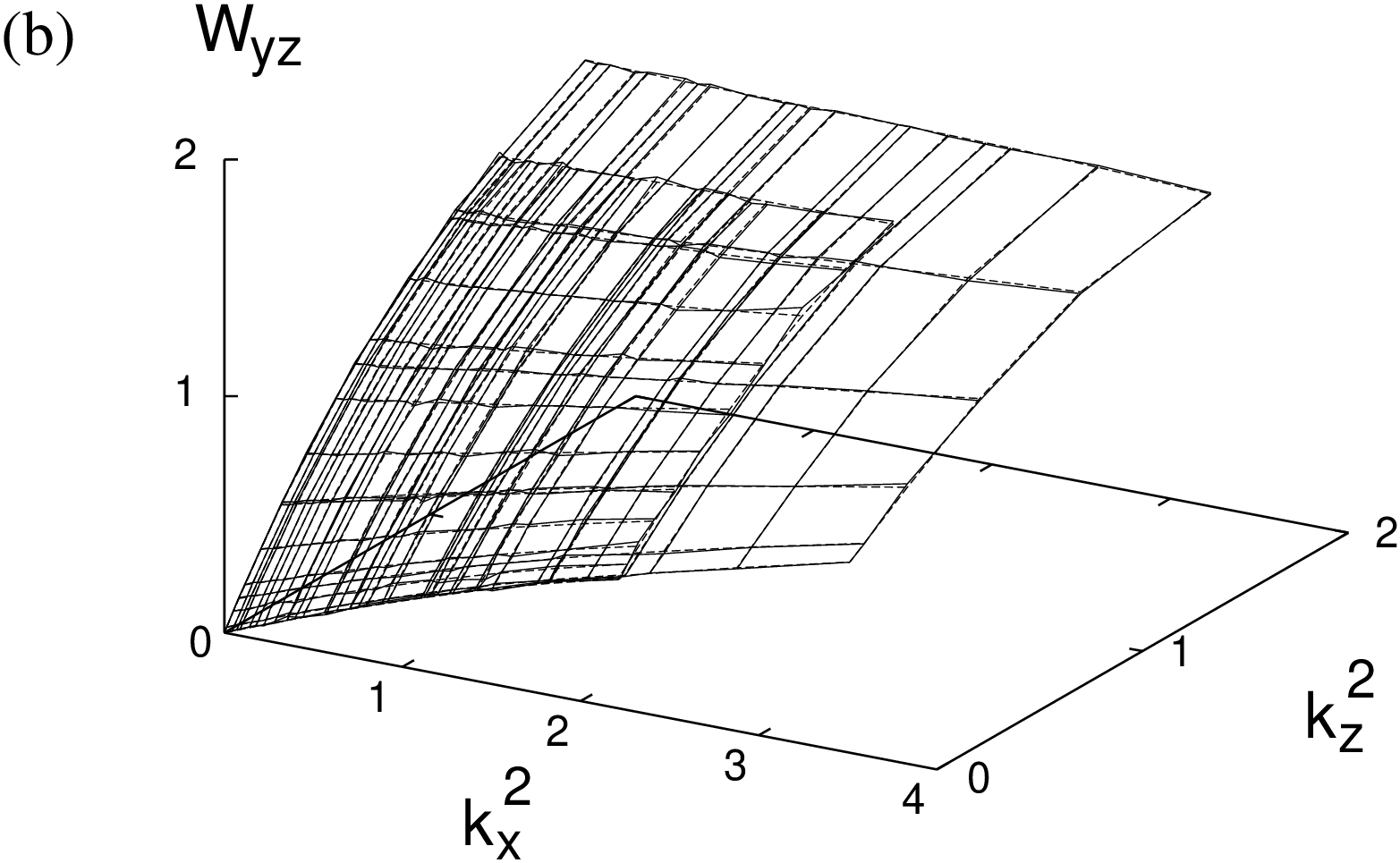}{63}{60}{
\vspace*{0.2cm}
\CCww
}
\end{figure}

\noindent
In our simulations, we obtained 1.13, which is in good agreement
with the theoretical value 1.12.

The results for the ${\cal W}_{\alpha z}(\kk)$ surfaces are shown in 
Fig.~\ref{fig:ww}. The data for the small system (coarse grid) 
match almost exactly those for the large system (fine grid). 
They were fitted to a fourth order polynomial in $k_x^2$ and $k_z^2$ 
(i.e., with highest order terms $k_x^8, k_x^6 k_z^2, \cdots, k_z^8$)
without a zeroth order term.
Higher orders were disregarded because the 4th order coefficients turned
out to be already very small. Normal equations and singular value
decomposition gave the same results. Fig.~\ref{fig:ww} demonstrates
that the fit is almost perfect. The leading coefficients give the
elastic constants, shown in table \ref{tab:el1}.
As expected for elongated molecules, one finds that $K_{33}$ is largest,
followed by $K_{11}$ and $K_{22}$.

Next we discuss the results for the pair correlation functions.
The spherical harmonics expansion coefficients of the pair distribution 
function $\rho^{(2)}$ were determined using~\cite{streett}
\begin{eqnarray}
\lefteqn{\rho^{(2)}_{l_1 m_1 l_2 m_2 l m}(r) = 4 \pi \: \varrho^2 \: g(r)}
\qquad \qquad \nonumber\\
\label{eq:streett}
&&
\langle 
\: Y^*_{l_1 m_1}(\uu_1) Y^*_{l_2 m_2}(\uu_2) Y^*_{l m}(\ru) \:
\rangle_{\delta r},
\end{eqnarray}
where $\langle \cdot \rangle_{\delta r}$ denotes the average over all molecules
in a shell $\delta r$ from $r$ to $r + \delta r$, and the function $g(r)$ is
the number of molecular centers at distance $r$ from a given molecular center,
divided by the number at the same distance in an ideal gas at the same density.
The calculation of these averages is very time consuming, since a great number
of coefficients has to be evaluated, and was therefore carried out in part on a
massively parallel machine. We have determined coefficients for values of
$l,l_i$ up to $l_{max}= 6$ in all systems, and for values up to $l_{max}=8$ 
in the smallest system. The bin size was $\delta r = 0.04$ and the cutoff 
distance $r_{max}$ was chosen to be 40\% of the box side $L$ in order to 
reduce boundary effects~\cite{pratt}. 

From the pair distribution function we calculated the total correlation 
function by inverting Eqn.~(\ref{eq:TCF_2}). The latter was then Fourier
transformed according to Eqn.~(\ref{eq:hankel}). There is a subtle problem 
here: due to the elasticity of the nematic phase, the total correlation 
function decays algebraically like $1/r$. This follows directly from
Eqn.~(\ref{eq:frank})~\cite{forster}. Before applying Eqn.~(\ref{eq:hankel}),
we thus fitted the simulation data points at the largest distances $r > r_0$
to a power law of the form $b/r$ and extrapolated $h(r)$ to infinity
\cite{footnote5}. The parameter $r_0$ was chosen to be $2.8, 4.0$ and $5.3$
in systems of $N=1000, 4000$ and $8000$ particles, respectively.

\begin{table}[t]
\CCtaba
\end{table}

\begin{figure}[t]
\noindent
\hspace*{-1cm}
\fig{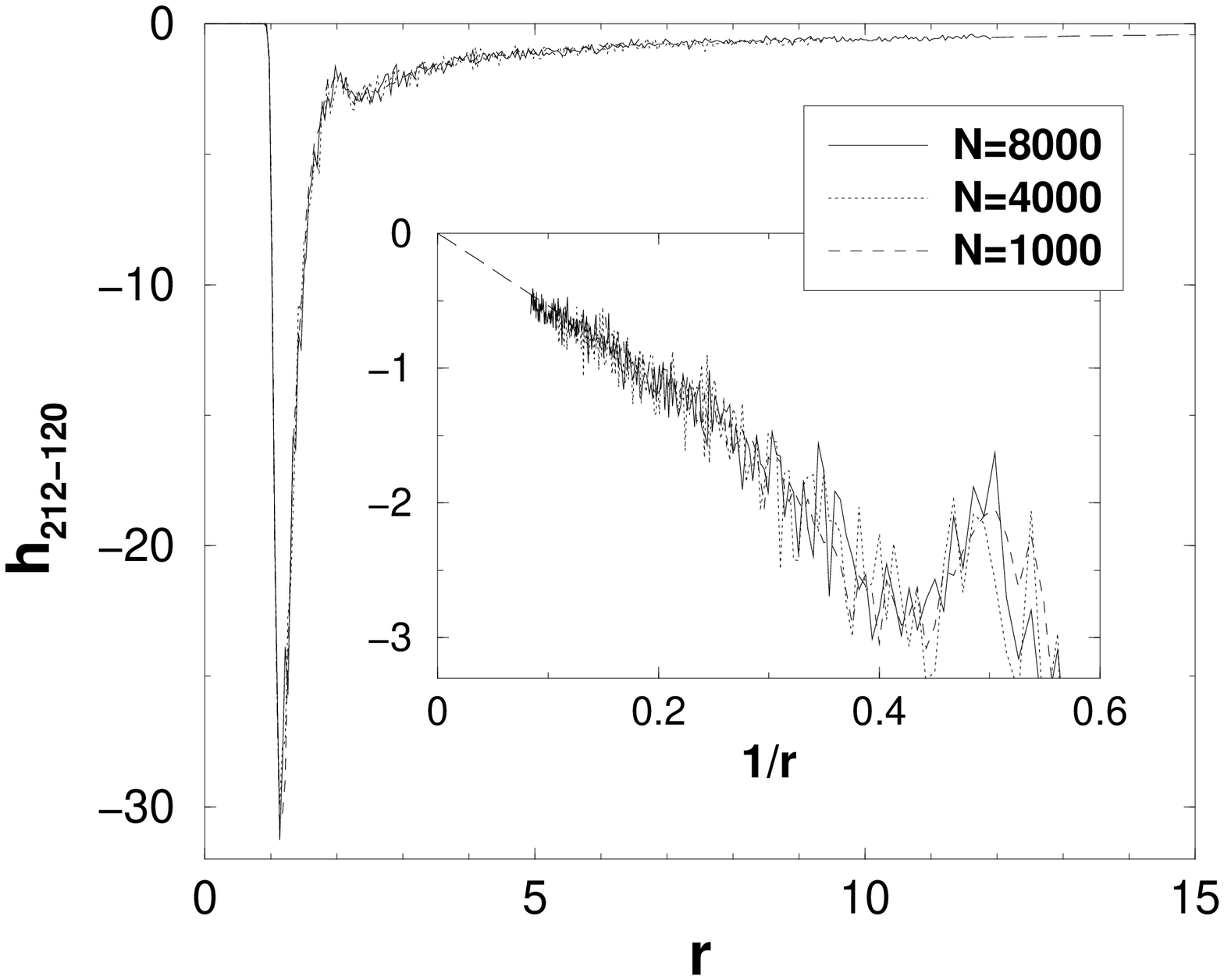}{80}{70}{
\vspace*{0cm}
\CCextrap
}

\end{figure}

It turned out that the long-range tail was quite pronounced for coefficients 
of $h$ with $m_1=\pm 1, m_2=\pm 1$, and almost negligible for the others. 
In Fig.~\ref{fig:extrap} we show an example of a coefficient with a 
pronounced long-range tail, the coefficient $h_{l_1 m_1 l_2 m_2 l m}(r)$ 
with $l_1=l_2=l=2$, $m_1=1$, $m_2=-1$ and $m=0$.
The data for different system sizes $N=1000$, $N=4000$, and $N=8000$ 
lie almost on top of each other, hence the form of $h(r)$ at
$r < r_{max}$ is not affected by noticeable finite size effects.
The dominating finite size problem comes from the uncertainty 
of the extrapolation, if the available range of $h(r)$ is too short.

The rest of the analysis was straightforward. From the coefficients of the 
total correlation function in Fourier space, $h_{l_1 m_1 l_2 m_2 l m}(k)$,
those of the DCF were obtained by solving the linear matrix equation
(\ref{eq:OZ_2}). Then we calculated the functions $C_{ii}(k)$ as defined
in Eqn.~(\ref{eq:cii}). According to Eqn.~(\ref{eq:ps_2}), the elastic 
constants $K_{ii}$ can be determined from the initial slopes in a plot 
of $C_{ii}(k)$ versus $k^2$. Data for $C_{ii}(k)$ are shown for different 
system sizes in Fig.~(\ref{fig:cii}). The points at zero wavevector 
$C_{ii}(0)$ were calculated using Eqn.~(\ref{eq:cond_2}). 
They fit nicely on the straight lines at $k\to 0$, hence the data are 
consistent with the requirement (\ref{eq:cond}) or (\ref{eq:cond_2}). 
This gave additional confidence in the quality of the analysis. 
The slopes of the straight lines yield the elastic constants.

The results are summarized in table \ref{tab:el2}. We have
calculated the DCF from the pair distribution function $\rho^{(2)}$
using an upper cutoff $l_{max}=2,4$ and $6$, respectively, in the 
matrix equations~(\ref{eq:TCF_2}) and (\ref{eq:OZ_2}). Already the lowest 
order calculation with $l_{max}=2$ gave elastic constants of the correct 
order of magnitude. Quantitatively reliable results were obtained with
$l_{max}\ge 6$: we checked in the smallest system that the results from
calculations with $l_{max}=6$ and $l_{max}=8$ do not differ
significantly.

Since the calculations with $l_{max}=8$ were very time consuming
(one has 1447 different expansion coefficients), we used $l_{max}=6$ in the
analyses of the larger systems (469 different expansion coefficients).

The results were the same for systems of size $N=1000,4000$ and 8000.
Furthermore, they were not affected by the presence of a director 
constraint: as mentioned in

\begin{figure}[t]

\noindent
\hspace*{-0.5cm}
\fig{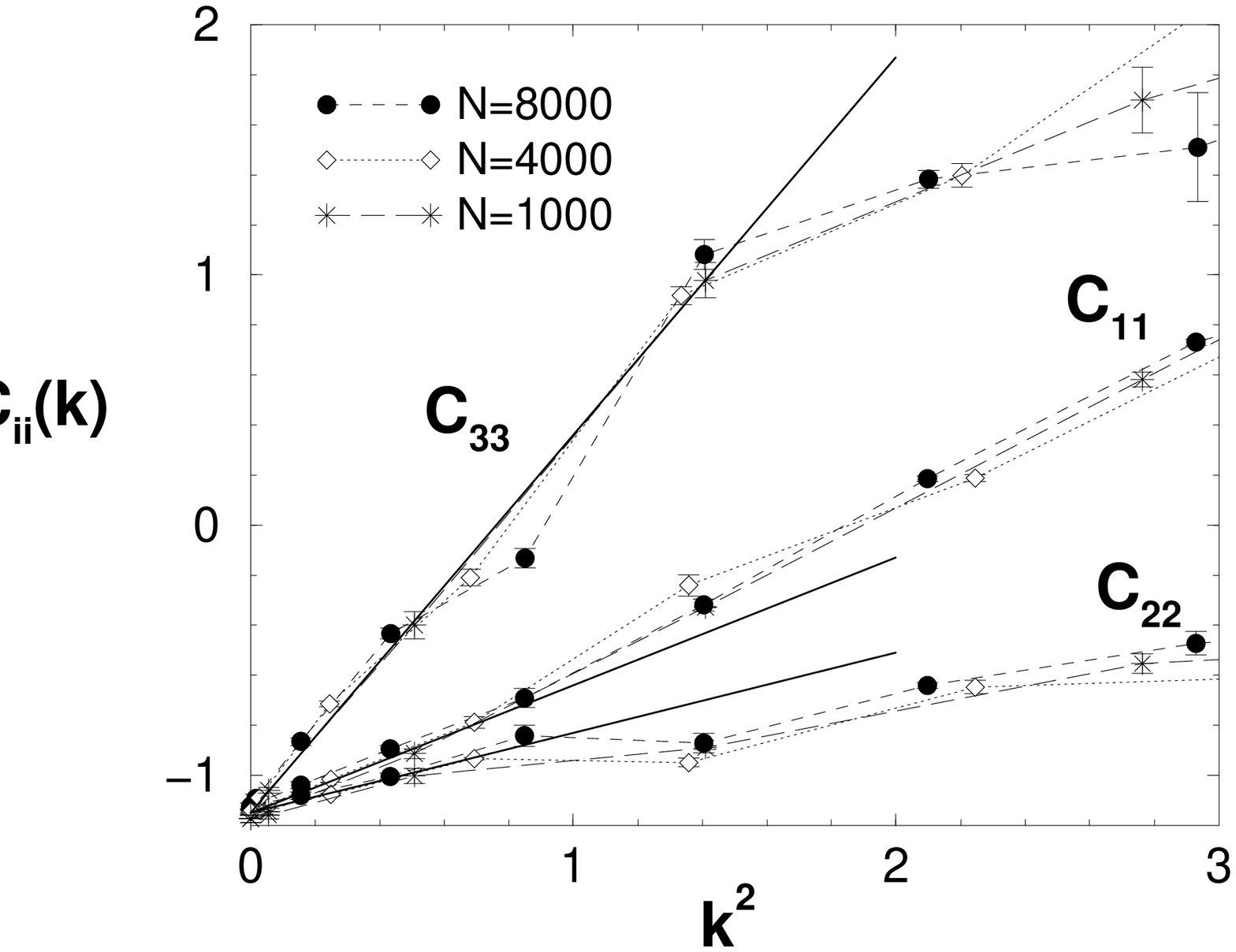}{75}{70}{
\vspace*{0cm}
\CCcii
}
\end{figure}

\vspace*{-0.2cm}
\begin{table}[t]
\CCtabb
\end{table}

\noindent
section \ref{sec:simulation}, the DCF was 
mostly calculated in unconstrained systems, but we also studied the 
DCF in one constrained system for comparison.

Finally, we compare the values of the elastic constants calculated by the 
DCF approach with those obtained from the order fluctuation analysis, 
shown in table \ref{tab:el1}. The values for $K_{11}$
and $K_{22}$ are identical for both methods. $K_{33}$ is slightly 
underestimated by the DCF analysis with $l_{max}=6$, but the result
increases with $l_{max}$, and agrees within the error with that of 
the order fluctuation analysis at $l_{max}=8$. 

One might ask how much the successive coefficients of the (correct) DCF 
contribute to the elastic constants. We found that the contribution 
of the coefficients with $l_1>4$, $l_2>4$ or $l>4$ is very small. 
If we include only terms up to $l,l_i = 4$ in Eqn.~(\ref{eq:cii}), 
we obtain $K_{11}=0.55$, $K_{22}=0.21$ and $K_{33}=1.51$ 
(in the largest system $N=8000$), which is very close to the final 
values quoted in table \ref{tab:el2}. However, we could not
push this analysis further. If we include only terms up to $l,l_i=2$,
the resulting $C_{ii}(k)$ are very concave and have no well-defined 
initial slope in a plot vs.~$k^2$. Hence the contributions of 
successive $l$s to the elastic constants cannot be distinguished.

\section{Summary and Conclusions}
\label{sec:summary}

We have presented a method which allows one to determine
without approximations the direct correlation functions in nematic liquid
crystals from computer simulations, and to calculate elastic constants 
on that basis according to the Poniewierski-Stecki theory~\cite{poniewierski1} 
(\ref{eq:ps1})$^-$(\ref{eq:ps3}). We have applied this method to a nematic 
fluid of soft ellipsoids. In the same system, the elastic constants were 
also determined by an established approach, the analysis of order 
tensor fluctuations. 

Our study represents a direct test of the Poniewierski-Stecki theory.
We found that the results obtained with the two methods agree well with 
each other. The Poniewierski-Stecki theory can thus be employed to 
calculate elastic constants, at least in our system, provided that the exact 
direct correlation functions are used in the equations. 

Hence we have established an alternative way of calculating elastic 
constants in nematic liquid crystals. As long as a simulation is performed
solely to determine elastic constants, the order tensor fluctuation approach 
is still more efficient: the statistical error of pair correlation functions 
must be quite small for a reliable DCF analysis, and the analysis is very 
time consuming. However, the DCF approach has the advantage of being 
straightforward; elastic constants can be computed from arbitrary bulk 
simulations, if the pair distribution functions are known with 
sufficient accuracy. Even the calculation of spatially varying elastic
constants, e.g., in the vicinity of surfaces, is conceivable. 

The direct correclation function is a central quantity in liquid state
theories. The study of direct correlation functions in the nematic phase 
is therefore interesting in its own right. We shall examine them in more 
detail and compare them to those in the isotropic phase in a 
forthcoming publication~\cite{phuong2}.

\section*{Acknowledgments}
 
We thank M.~P.~Allen and H.~Lange for fruitful interactions, and
P.~Teixeira for helpful comments on the manuscript. The simulations
and the analyses were run in part on a CRAY T3E of the HLRZ in J\"ulich. NHP
and GG received financial support from the German Science Foundation (DFG). The
parallel MD program \textsc{gbmega} used in this work was originally developed
by the EPSRC Complex Fluids Consortium, UK.


\begin{thebibliography}{99}
\bibitem{footnote1} There are differences, however:
  In Refs.~\onlinecite{singh1}$^-$\cite{paulo}, 
  the direct correlation function in the equations is that of an 
  isotropic reference state, whereas in the original 
  Poniewierski-Stecki equations it is that of the nematic phase.
  The expressions derived by Singh and Singh~\onlinecite{singh1,singh2} 
  differ from the Poniewierski-Stecki equations, but they can be cast 
  in the form (\ref{eq:ps1})--(\ref{eq:ps3}) if one exploits the 
  symmetry of the reference state.
\bibitem{footnote2} The pressure was determined from the simulation data.
  The temperature was a fixed parameter in the Monte Carlo simulations. 
  The molecular dynamics simulations were carried out in the microcanonical 
  ensemble with the total energy adjusted such that the temperature had
  the desired value. Statistical errors are in the third decimal place.
\bibitem{footnote3} In our scaling, the time unit is 
  $\sqrt{m_0/\epsilon_0} \sigma_0$.
\bibitem{footnote4} The choice of an elongated box was motivated by
  the fact that $K_{33}$ is about twice as large as $K_{11}$ and
  $K_{22}$. Eqns.~(\ref{eq:qfluc1}), (\ref{eq:qfluc2}) or (\ref{eq:ww1}), 
  (\ref{eq:ww2}) reveal that it is of advantage to work on a $\kk$-grid, 
  which is finer in the $z$-direction than in the other two.
\bibitem{footnote5} For a few coefficients, it was also necessary to shift 
  the data by a constant. See Ref.~\onlinecite{phuong2} for a more detailed
  presentation of the data analysis.
\end{thebibliography}
\end{document}